\documentclass[nofootinbib,superscriptaddress,showpacs,showkeys]{revtex4}

\usepackage{graphicx,epsfig}
\usepackage{amsfonts,amsmath,amssymb,amsthm,amscd}
\usepackage[T2A]{fontenc}
\usepackage{color}

\begin{document}

\title{Relativistic spherical plasma waves}

\author{S. S. Bulanov}
\affiliation{University of California, Berkeley, California 94720, USA}

\author{A. Maksimchuk}
\affiliation{Center for Ultrafast Optical Science,
University of Michigan, Ann Arbor, Michigan 48109, USA}

\author{C. B. Schroeder}
\affiliation{Lawrence Berkeley National Laboratory, Berkeley, California 94720, USA}

\author{A. G. Zhidkov}
\affiliation{Central Research Institute of Electric Power Industry,
Yokosuka, Kanagawa 240-0196, Japan}

\author{E. Esarey}
\affiliation{Lawrence Berkeley National Laboratory, Berkeley, California 94720, USA}

\author{W. P. Leemans}
\affiliation{University of California, Berkeley, California 94720, USA}
\affiliation{Lawrence Berkeley National Laboratory, Berkeley, California 94720, USA}

\begin{abstract}
Tightly focused laser pulses as they diverge or converge in underdense plasma can generate wake waves, having local structures that are spherical waves. Here we report on theoretical study of  relativistic spherical wake waves and their properties, including wave breaking. These waves may be suitable as particle injectors or as flying mirrors that both reflect and focus radiation, enabling unique X-ray sources and nonlinear QED phenomena.  
\end{abstract}

\maketitle

{\noindent} The study of nonlinear oscillations in plasma \cite{AkhiezerPolovin, Dawson} is one of the great successes of plasma physics, both from  understanding the underlying processes and from applications. Nonlinear waves, driven by intense lasers, offer a unique possibility to accelerate charged particles \cite{wake}, providing the longitudinal electric fields of the unprecedented strength (see \cite{review} and references cited therein). However the implementation of these schemes had to wait until the invention of Chirped Pulse Amplification \cite{CPA}, which allowed for the generation of extremely short electromagnetic (EM) pulses, and until the understanding was gained of the mechanisms of injection of electron bunches into the plasma wave accelerating fields \cite{SVB}. Subsequently the process of electron acceleration by nonlinear plasma wake waves was demonstrated in a number of pioneering experiments \cite{Nature}.

The nonlinear waves in plasmas can be used not only for the acceleration of charged particles, but they also can be used to generate ultra-bright X- and $\gamma$-ray radiation via the mechanism that is usually referred to as a Flying Mirror (FM) \cite{LI}. This mechanism makes use of the fact that the nonlinear waves in the wavebreaking regime are characterized by the formation of "spikes" in the electron density, where the density tends to infinity. Such modulations of the electron density moving with the relativistic velocity and acting as a FM are able to reflect a counterpropagating EM wave in the form of high-frequency radiation \cite{LI-exp}.

In this Letter we describe the basic properties of nonlinear oscillations in plasma, extending the analysis done by Dawson for the case of nonrelativistic spherical waves \cite{Dawson} to the relativistic regime and discuss their applications. We notice that such waves demonstrate behavior typical for waves with continuos spectra, i.e., the phase mixing and consequent wave breaking, which can be used for electron acceleration and FM generation. As was shown in Ref. \cite{SVB} the breaking of a plasma wave provides means of injection of an electron bunch into the wakefield for subsequent acceleration. For spherical plasma waves, phase mixing is an intrinsic property and wavebreaking takes place at a specific distance from the center of the sphere, providing a way for controllable injection. The spherical wave traveling towards the center of the sphere in the wavebreaking regime is a good candidate for the FM. The spherical flying mirror (SFM) compresses and tightly focuses the incoming radiation, thus achieving the highest intensification factor among all the FM schemes.        

Assuming that ions are at rest, we utilize the model of a collisionless cold plasma, governed by Maxwell equations and by hydrodynamics equations of an electron fluid in the fixed ion background 
with homogeneous ion density $n_0$. We consider the case when all the variables that characterize the fields and the plasma depend only on the radius $r$ and time $t$:   
\begin{eqnarray} \label{cont}
\partial_t n+r^{-2}\partial_r(r^2nv)=0, \\
\label{motion1}
\partial_t p+v\partial_r p=-E, \\
\label{motion2}
v=p(1+p^2)^{-1/2},\\
\label{Maxwell}
r^{-2}\partial_r(r^2 E)=1-n,
\end{eqnarray}
where we normalized the time and space coordinate to $\omega_{pe}^{-1}$ and $c\omega_{pe}^{-1}$, the electron density $n$ to $n_0$, the electron momentum $p$ and velocity $v$ to $m_e c$ and $c$, and the electric field $E$ to $m_e \omega_{pe} c/e$, respectively, with $\omega_{pe}=(4\pi n_0 e^2/m_e)^{1/2}$ being the plasma frequency.

We perform transformation from the Euler coordinates $(r,t)$ to the Lagrange coordinates $(r_0, t)$ related to each other by $r=r_0+\xi(r_0,t)$, where $r_0$ is the initial position of the electron fluid element and $\xi(r_0,t)$ is its displacement. Solving the continuity equation (\ref{cont}) we obtain for the electron density:
\begin{equation}
n=n_0 \left[J(r_0,t)\right]^{-1},
\end{equation}
where $J(r_0,t)=(1+\partial\xi/\partial r_0)(r_0+\xi)^2/r_0^2$ is the Jacobian of the transformation from the Lagrange to the Euler coordinates. The transformation has a singularity at the point where	the	Jacobian vanishes $J(r_0,t)\rightarrow 0$.	 This singularity occurs either when $\xi=-r_0$ or for $\partial\xi/\partial r_0=-1$, which is known as a condition of the wave breaking, when the electron density and the gradient of the electron velocity become infinite.
 
The time dependence of the electron displacement, $\xi=\xi(r_0,t)$, is determined from Eqs. (\ref{motion1}) - (\ref{Maxwell}), which in the Lagrange coordinates take the form:
\begin{eqnarray} 
\label{motion-p}
\partial_t p=-E, \\
\label{motion-xi}
\partial_t\xi=p(1+p^2)^{-1/2}, \\
\label{E-Lagr}
\partial_t [(r_0+\xi)^2 E]=(r_0+\xi)^2 \partial_t\xi.
\end{eqnarray}
The solution of Eq. (\ref{E-Lagr}) yields the expression for the electric field $E$ 
\begin{equation}\label{field}
E=\frac{(r_0+\xi)^3- r_0^3}{3(r_0+\xi)^2}.
\end{equation}
Using this we find the integral of motion:
\begin{equation}
\label{integral}
(1+p^2)^{1/2}+\frac{(r_0+\xi)^3+2 r_0^2(r_0-\xi)}{6(r_0+\xi)}=h,
\end{equation}
where the constant $h$ marks the trajectory in the phase plane $(p,r_0+\xi)$. Equation (\ref{integral}) indicates that the oscillations can never reach the center of the sphere due to the singularity at $\xi=-r_0$. This also can be seen in Fig. 1a where we present the phase plane of electron motion, $(p,r_0+\xi)$, which illustrates the nonlinear character of plasma oscillations. 

\begin{figure}[tbp]
\epsfxsize7.7cm\epsffile{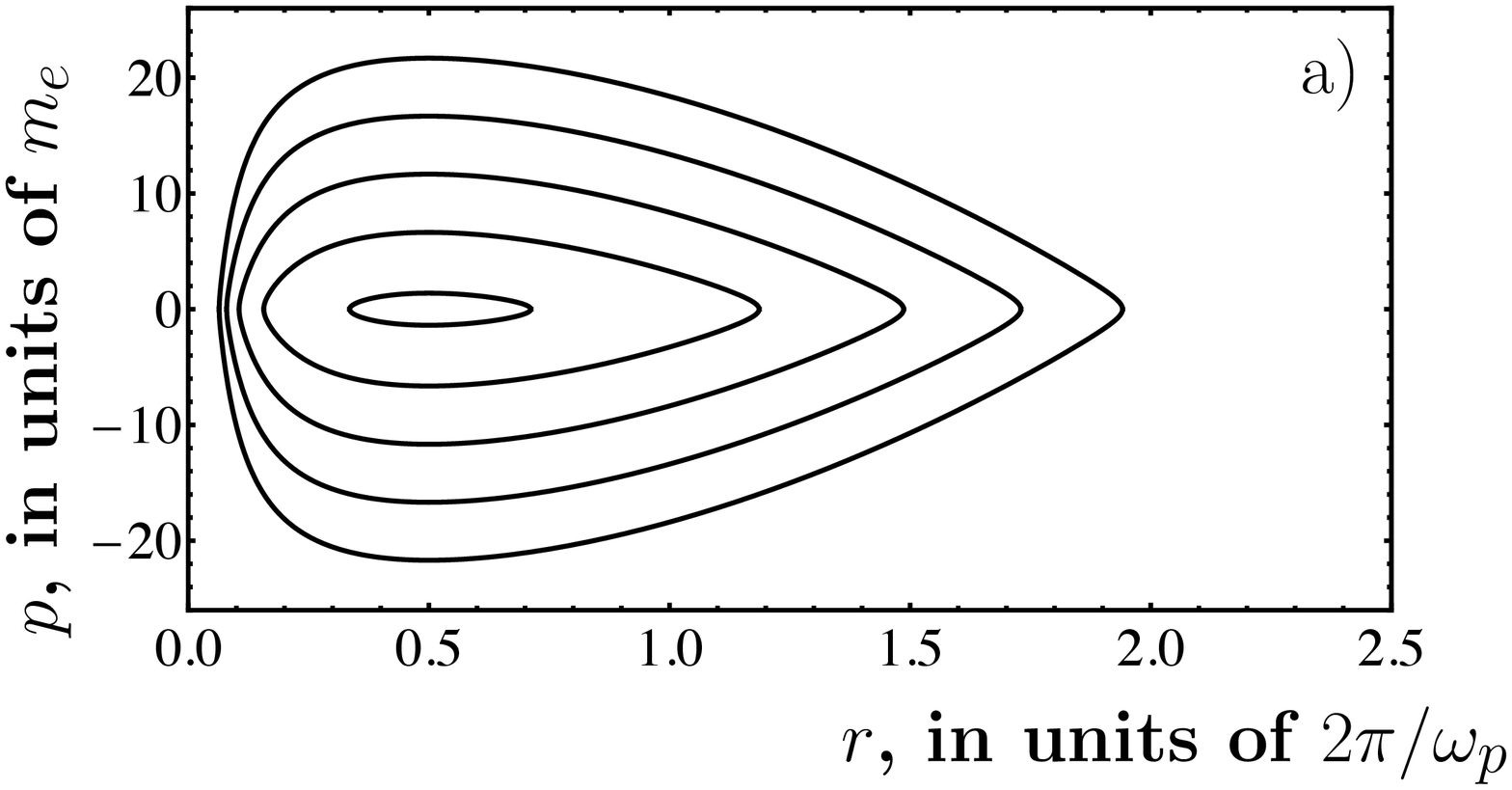} \epsfxsize7cm\epsffile{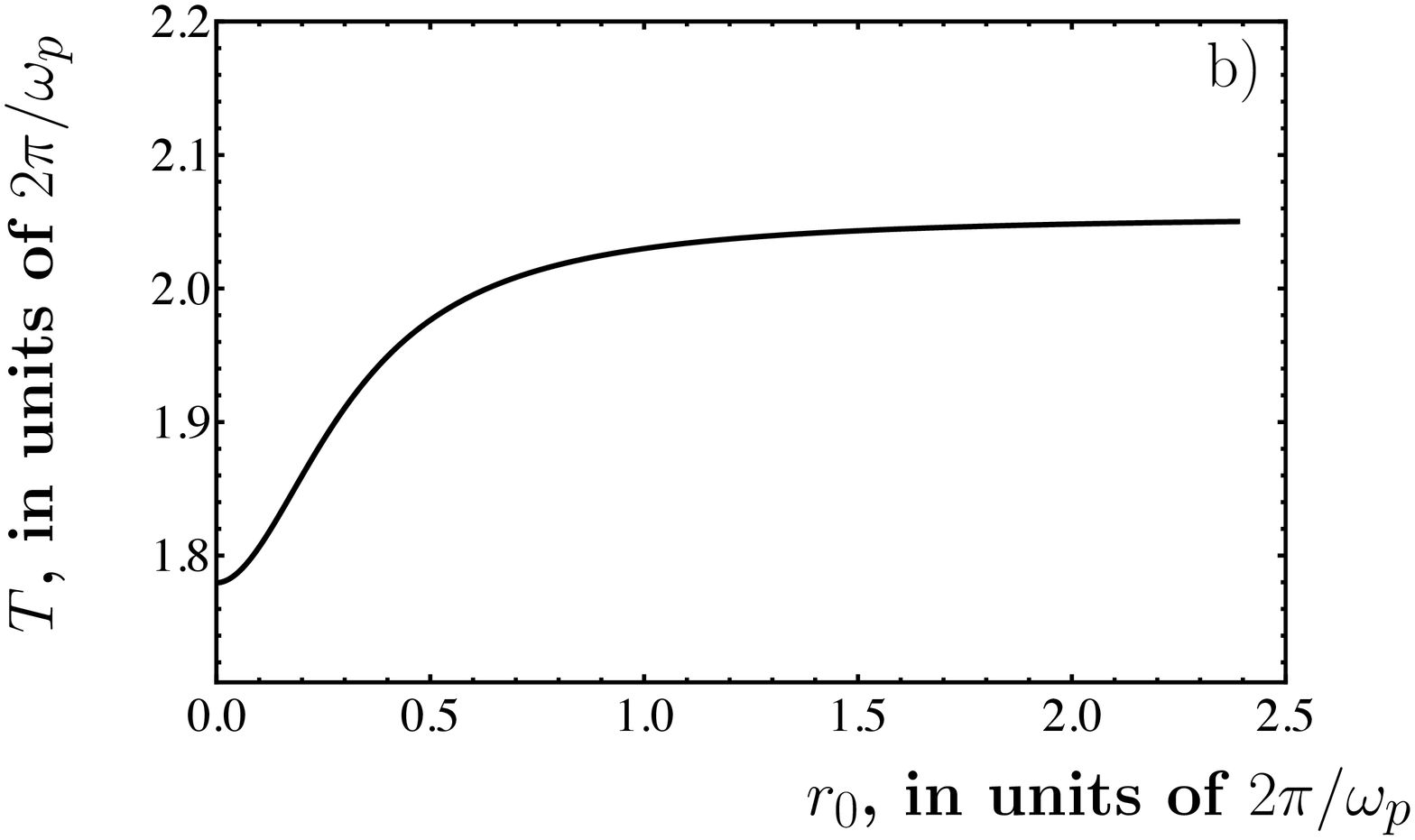} 
\caption{ (a) Phase plane $(p,r_0+\xi)$ for $r_0=2$ and $h=5,10,15,20,25$. 
(b) Dependence of the oscillation period on the radius, $p_0=5$.}
\end{figure}

Using the integral of motion given by Eq. (\ref{integral}) we obtain the solution of Eq. (\ref{motion-xi}) in quadrature:
\begin{equation}\label{sol1}
t=\int\limits_0^\xi d\xi^\prime \frac{\left[p(\xi^\prime)^2+1\right]^{1/2}}{p(\xi^\prime)}.
\end{equation}
The period of oscillations can be determined from this equation by integrating over the entire oscillation cycle, i.e. by choosing the integration limits to be equal to $\xi_{min}$ and $\xi_{max}$, which are the minimum and maximum values of coordinate $\xi$ and determined from the $p=0$ condition in Eq. (\ref{integral}): 
\begin{equation}\label{xi_min}
\xi_{min,max}=2\left[2(h-1)+r_0^2/3\right]^{1/2}\cos\left[\left(\Theta\mp\pi\right)/3\right],
\end{equation}
where $\Theta=\arctan\left\{1-r_0^{-6}\left[2(h-1)+r_0^2/3\right]^3\right\}$ and minus or plus corresponds to $\xi_{min}$ or $\xi_{max}$, respectively. For $p_0\gg 1$ the period of oscillations $T$ is
\begin{equation}
T=2(6p_0)^{1/2},
\end{equation}
where $p_0$ is maximum amplitude of the momentum oscillation. In a weak nonlinearity limit we can solve the equations of motion. Assuming  $\xi\ll r_0$, and $\partial_t \xi \ll 1$, we obtain from Eq. (\ref{motion-xi})
\begin{equation}\label{linearized}
\partial_{tt}\xi+\xi=\xi^2/r_0
-(4\xi^3 /3r_0^2)+(3/2)\xi\left(\partial_t\xi\right)^2.
\end{equation}
It can be seen from the form of this equation that the nonlinear terms are of two types. The first type (the first two terms in the r.h.s) is due to the spherical geometry of the oscillations, while the second one (the third term) is due to the relativistic effects \cite{Liu}.
 
This equation can be solved using perturbations theory by expanding in series the displacement $\xi$ and the oscillation frequency $\omega$: $\xi=\xi_0+\xi_1+...$ and $\omega=1+\omega_1+...$. The zeroth order solution to Eq. (\ref{linearized}) is $\xi_0=\kappa \cos t$ with $\kappa$ being the oscillation amplitude, which in general depends on the coordinate $r_0$. The frequency is determined from the condition that there are no secular resonant terms in the equations for $\xi_1,\xi_2,...$. We find that the oscillation frequency depends on the coordinate and amplitude as
\begin{equation}
\omega=1+\kappa^2/(12r_0^2)-3\kappa^2/16.
\end{equation} 
The frequency grows while the coordinate $r_0$ decreases in accordance with the result of numerical calculation of oscillation period based on Eqs. (\ref{sol1}) and (\ref{xi_min}), which is plotted in Fig. 1b.

The frequency dependence on the coordinate means that the spherical plasma wave has a continuos spectrum. Considering the evolution of such waves we assume they have a form $a \exp(i\psi)$ with the eikonal $\psi(r_0,t)$. Then the wavenumber and frequency are expressed via the eikonal derivatives,  $k=\partial_{r_0}\psi$ and $\omega=-\partial_t\psi$. By virtue of crossdifferentiation we have $\partial_t k +\partial_{r_0}\omega=\partial_t \partial_{r_0}\psi-\partial_{r_0}\partial_t \psi=0$. This yields $\partial_t k =-\partial_{r_0} \omega$. Since $\partial_{r_0} \omega$ does not depend on time we obtain $k=k_0-\partial_{r_0} \omega ~ t$. If the wave was initially converging, $k_0<0$ and $(\partial \omega/\partial r_0)<0$, or for initially diverging wave $k_0>0$ and $\partial_{r_0}  \omega >0$, then at the time, $t=t_{turn}$,where
\begin{equation}
t_{turn}=k_0/\partial_{r_0}  \omega,
\end{equation} 
the wave will stop and then go in the opposite direction. The sign of the gradient of the nonlinear plasma wave oscillation frequency is determined by the 3D geometry factor or/and by the dependence of the wave amplitude $\kappa$ on the coordinate.

\begin{figure}[tbp]
\epsfxsize10cm\epsffile{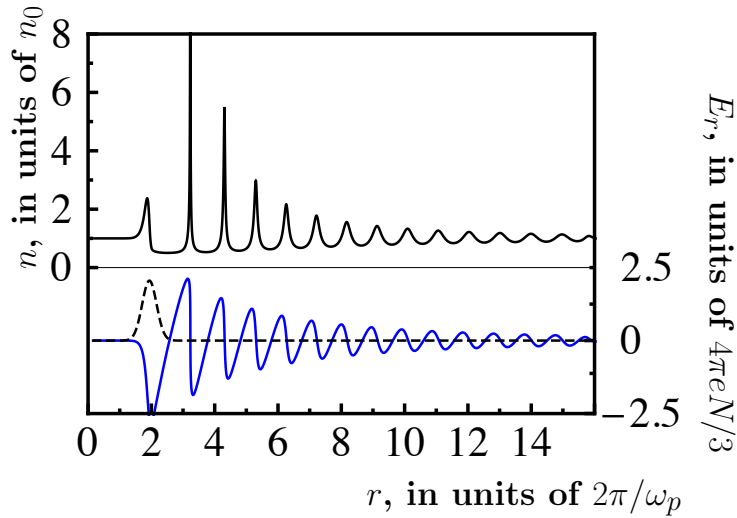} 
\caption{ The dependences of density and longitudinal electric field on the radius 
in the case of the wake generation by the laser pulse. The field of the driving pulse is showed by a dashed line.}
\end{figure}

The time of the wave breaking can be estimated following the results of Ref. \cite{Brantov}. The displacement $\xi$ can be represented as a superposition of oscillations that are the harmonics of plasma frequency, which depends on the distance from the center of the sphere, $r_0$:
\begin{equation}
\xi(r_0,t)=\sum_m\xi_m(r_0)\exp\left[im\omega(r_0)t\right].
\end{equation}
Differentiating $\xi(r_0,t)$ with respect to  $r_0$ and substituting it into the wavebreaking condition $1+\partial \xi/\partial r_0=0$, we obtain for the breaking time:
\begin{equation}
t_{br}=-\omega/\left(\partial_{r_0} \omega~\partial_t\xi\right).
\end{equation}
Assuming that for a converging plasma wave near the breaking point the frequency can be approximated as $\sim r_0^{-1}$, then for $\partial_t \xi\rightarrow 1$ we have $t_{br}\sim r_0$ and for $\partial_t \xi\ll 1$ the wavebreak time is $t_{br}\sim r_0 (\partial_t \xi)^{-1}$. This means that in the relativistic case the first period of the wave breaks at $r_0\sim 1$, i.e., at the distance of about a plasma wavelength from the center. This behavior of nonlinear plasma waves can provide a mechanism of controllable injection of electrons into a laser wakefield accelerator.

Now we consider a spherical wake wave driven by an ultrashort EM pulse. We describe the pulse by introducing the ponderomotive force \cite{review} in the r.h.s of equations of motion as $f_p=-\partial_r\gamma$, where $\gamma=(1+p^2+a^2)^{1/2}$, $a$ is the vector potential of the laser pulse and the pulse electric field has the form $\exp(-(r_0-t)^2) Erf(r_0)/r_0$, and taking into account that $\partial_r=\partial_{r_0}/\partial_{r_0} r$. However if the plasma wave does not break inside the EM pulse and the oscillations are far from the center of the sphere we have  $\partial_{r_0} r \approx 1$. The results of the numerical solution of the motion equations in the presence of the ponderomotive force of this form are presented in Fig. 2 for the converging pulse. Pronounced spikes in the electron density can be seen, meaning that the wake wave entered the wave breaking regime, which can be utilized in the relativistic SFM scheme. Since the $\gamma$-factor of a breaking plasma wave is determined by the a group velocity of the driving laser pulse ($\gamma_g\approx\omega(a/2)^{1/2}$) and the amplitude of the laser pulse increases as $1/r$ as it propagates towards the focus, it follows that $\gamma_{SFM}\approx2^{-2/3}\omega^{4/3}$ at $r_M\approx a_{0}2^{-1/6}\omega^{-2/3}$.    
     
In what follows the results of 2D PIC simulations of intense tightly focused laser pulse interaction with underdense plasma are shown. The simulations were performed using the REMP (Relativistic ElectroMagnetic Particle) code \cite{Esirkepov_code}. The typical run utilized a simulation box with mesh grid size of $\lambda/20$. 

\begin{figure}[tbp]
\epsfxsize7cm\epsffile{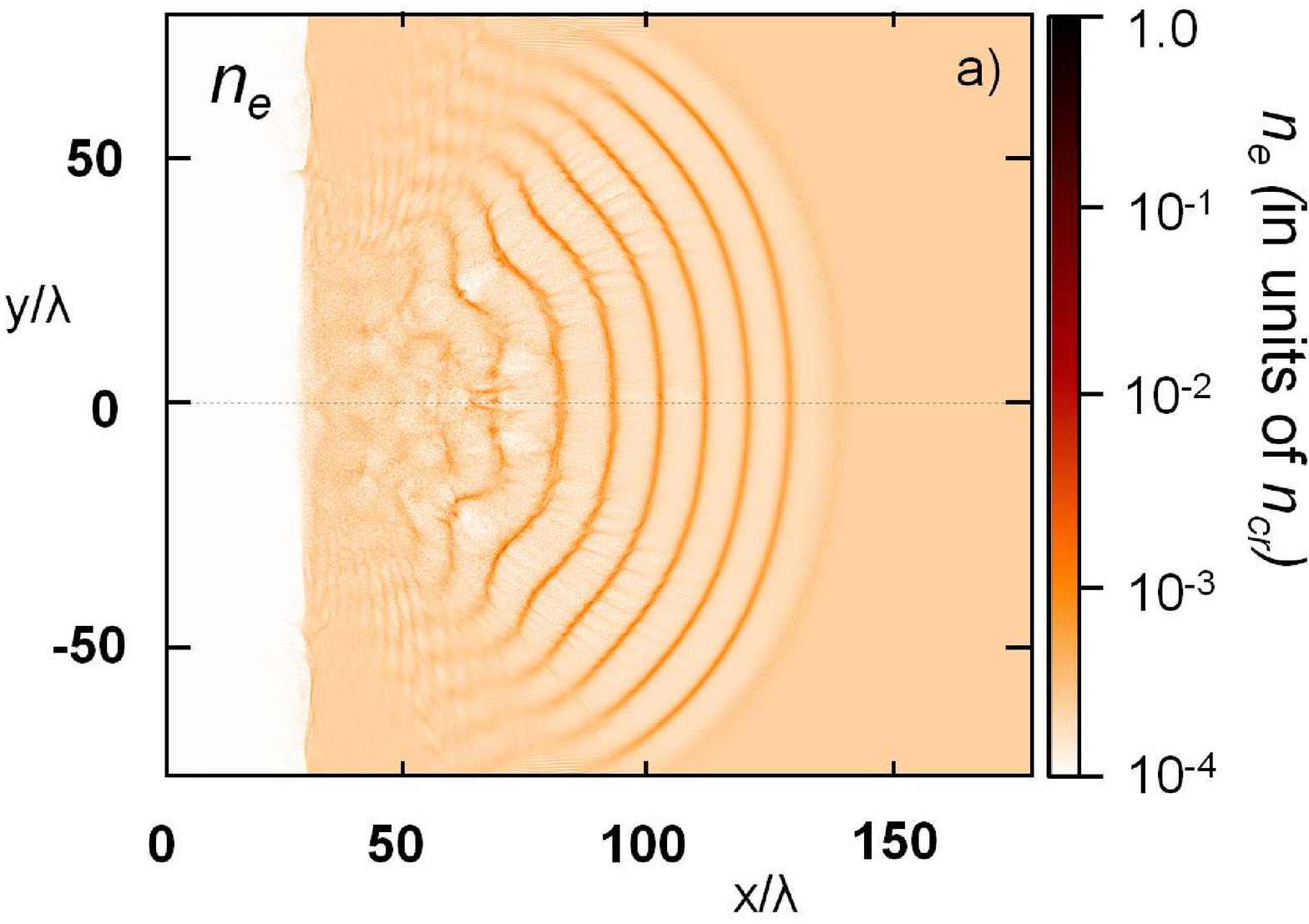} \epsfxsize7cm\epsffile{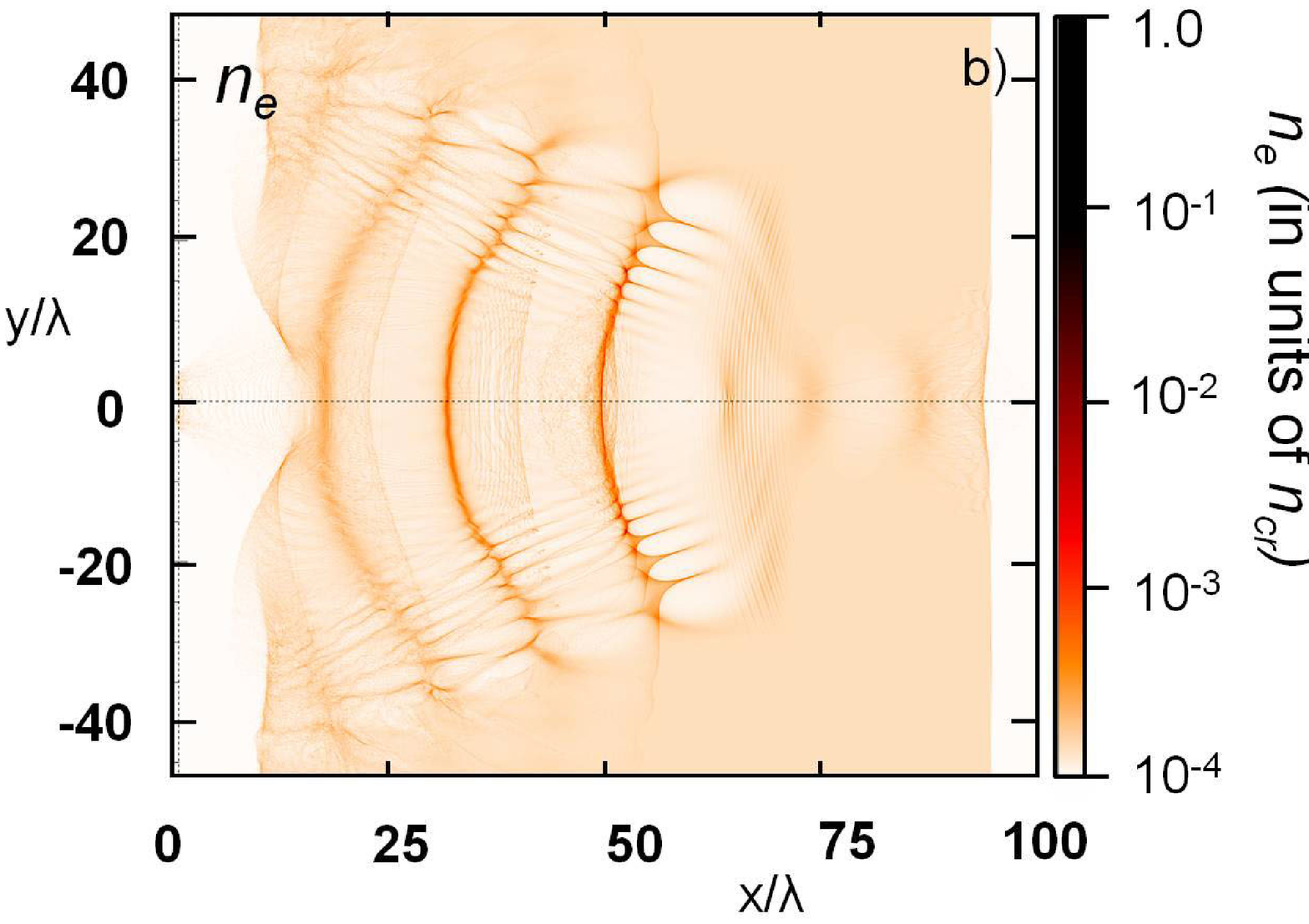} 
\caption{(Color on-line) Diverging and converging spherical wake wave: density profiles}
\end{figure}

In Fig. 3 the distributions of electron density illustrating the diverging (3a) and converging (3b) plasma wake waves are presented. In the diverging case a linearly polarized (out of the simulation plane) laser pulse with duration of $10\lambda/c$, introduced at the left boundary with $a=5$ and the width of $20\lambda$ was focused at $x=20\lambda$, resulting in a focal spot of about $3\lambda$. It interacted with a $150\lambda\times 160\lambda$ hydrogen plasma with the density $n/n_{cr}=1.56\times 10^{-2}$. The laser pulse drives the diverging plasma wave. In the converging case a linearly polarized (in the simulation plane) laser pulse with $a=2$, duration of $5\lambda/c$  and width of $50\lambda$ at the left boundary was focused at $x=150\lambda$. It interacted with a $85\lambda\times 110\lambda$ hydrogen plasma with the density $ n/n_{cr}=5.6\times 10^{-3}$. The converging plasma wave was produced, which breaks near the focus in accordance with the theoretical estimates.  
 
This breaking, converging wave acting as a SFM can not only up-shift the frequency but also can focus the reflected radiation tightly, further increasing its intensity.  In order to illustrate this phenomenon  computer simulations were carried out of a counterpropagating laser pulse interacting with such converging SFM. The linearly polarized (out of the simulation plane) counterpropagating laser pulse originated at the right boundary with $a=2$, $\lambda_R=4\lambda$, width of $20\lambda$ and duration $5\lambda/c$.  In Fig. 4 the z-component of the electric field is shown. The reflected pulse is focused into a tiny spot and its frequency is substantially increased. 

\begin{figure}[tbp]
\epsfxsize7cm\epsffile{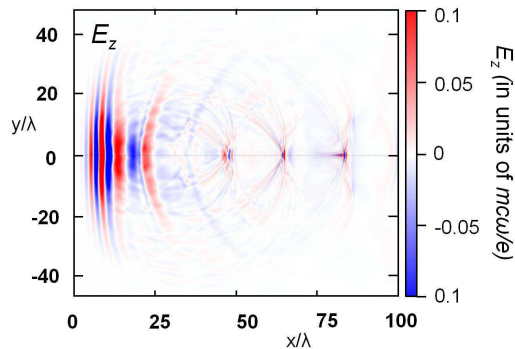}
\caption{(Color on-line) The reflection of a EM pulse from a
converging wake wave.}
\end{figure}

The typical intensity scaling is $I\sim \gamma^3 I_0$ when reflecting from a conventional (plane wave) FM \cite{LI}, where $I$ and $I_0$ are the intensities of the reflected and incident pulses respectively. However in our case the situation with a SFM is different since it is able not only to reflect but also to tightly focus the incoming radiation into a focal volume that is $64\gamma^6$ smaller than that of the incident wave. Assuming total reflection of the incident pulse the reflected pulse intensity scales as $I=256\gamma^8 I_0$. This is the maximum possible intensification factor assuming a reflection coefficient $R=1$. According to the results of Ref. \cite{LI}, the reflection coefficient $R\approx(\omega_d/\omega_s)^2/2\gamma^3$ at wave break, where $\omega_d$ and $\omega_s$ are the frequencies of the driver and source laser pulses, respectively. If $\omega_d=\omega_s$ then we have
\begin{equation} 
I=128\gamma^5 I_0.
\end{equation} 
This intensification factor exceeds all other obtained previously in different schemes using a FM \cite{LI, others}. 

Highly efficient SFM can in principle enable the study of non-linear effects of Quantum Electrodynamics (QED) with present day laser systems, and SFM can possibly lead to ultra-bright sources of X- and $\gamma$-rays. For $\gamma=15$ and 1 J energy of 1 $\mu$m wavelength  laser pulse reflected by a SFM will give rise to $10^{16}$ several keV photons. For higher intensities the energy of the emitted photons becomes of the order of the particle energy, so here the quantum recoil effects should be taken into account. Also at this point a process that has no classical analog appears: $e^+e^-$ pair production by a photon in a strong EM field. These non-linear QED processes paired together can lead to the electron-positron avalanche to occur \cite{BellKirk}. The parameters that characterize the probability of these processes $\chi_e=\sqrt{(F_{\mu\nu}p_\nu)^2}/m_e E_S$ and $\chi_\gamma=\sqrt{(F_{\mu\nu}\hbar k_\nu)^2}/m_e E_s$ should be of the order of unity for the avalanche to start. Our estimates show that the pulse with intensity of $I_0\sim 10^{22}$ W/cm$^2$ should be intensified by a SFM with $\gamma=10$ for this effect to be seen. Here $E_S=1.32\times 10^{16}$ V/m is the Schwinger field \cite{H&E}, $F_{\mu\nu}$ is the tensor of the EM field and $p_\nu$, $k_\nu$ are the 4-momenta of the electron and photon respectively. The SFM scheme will also enable the study of the Schwinger process \cite{H&E}, i.e., the production of $e^+e^-$ pairs from vacuum under the action of strong EM field. The estimates based on \cite{NarozhnyBulanovMurPopov} show that the intensity of $10^{27}$ W/cm$^2$ of the reflected pulse will be enough to observe the pair production. Such intensity will require $I_0\sim 10^{20}$ W/cm$^2$ and SFM with $\gamma\approx 10$. 

In conclusion, in this letter we studied the basic properties of the spherical oscillations in plasma. The produced spherical plasma waves demonstrate the behavior typical for waves with continuos spectra, i.e the phase mixing and consequent wave breaking. Since the breaking is due to the geometry of the oscillations, where and when it happens is determined by the properties of a plasma wave. Thus this wave can be used for controllable injection of electrons in the accelerating field of LWFA. In addition, the converging spherical wave may be suitable for a flying mirror, which both reflects and focuses incident radiation, enabling unique X-ray sources with applications to nonlinear QED. 

We thank G. Mourou and S. Wilks for discussions and T. Zh. Esirkepov for providing REMP code for simulations. We appreciate support from the NSF under Grant No. PHY-0935197 and through the FOCUS Center at the University of Michigan, the US DOE under Contract No. DE-AC02-05CH11231.


\begin{thebibliography}{99}
\bibitem{AkhiezerPolovin} A. I. Akhiezer and R. V. Polovin, Sov Phys. JETP \textbf{30}, 915 (1956).
\bibitem{Dawson} J. Dawson, Phys. Rev. \textbf{113}, 383 (1959).
\bibitem{wake} T. Tajima and J. Dawson, Phys. Rev. Lett. \textbf{43}, 267 (1979).
\bibitem{review}  E. Esarey, \textit{et al.},  Rev. Mod. Phys. \textbf{81}, 1229 (2009).
\bibitem{CPA} D. Strickland and G. Mourou, Opt. Commun. \textbf{56}, 219 (1985).
\bibitem{SVB} S. V. Bulanov, \textit{et al.}, JETP Lett. 53, 565 (1991);
D. Umstadter, \textit{et al.}, Phys. Rev. Lett. {\bf 76}, 2073 (1996);
S.V. Bulanov, \textit{et al., ibid.} {\bf 78}, 4205 (1997);
 E. Esarey, \textit{et al., ibid.} {\bf 79}, 2682 (1997); 
 S.V. Bulanov, \textit{et al.}, Phys. Rev. E {\bf 58}, R5257 (1998);
C. B. Schroeder, \textit{et al., ibid.} {\bf 59}, 6037 (1999); 
E. Esarey, \textit{et al.}, Phys. Plasmas {\bf 6}, 2262 (1999);
A. Pukhov and J. Meyer-ter-Vehn, Appl. Phys. B {\bf 74}, 355 (2002);
H. Suk, \textit{et al.}, Phys. Rev. Lett. {\bf 86}, 1011 (2001); 
A. Zhidkov, \textit{et al.}, Phys. Plasmas {\bf 11}, 5379 (2004); 
T. -Y. Chien, \textit{et al.}, Phys. Rev. Lett. {\bf 94}, 115003 (2005);
J. Faure, \textit{et al.}, Nature {\bf 444}, 737 (2006);
 C. G. R. Geddes, \textit{et al.}, Phys. Rev. Lett. {\bf 100}, 215004 (2008);
 H. Kotaki, \textit{et al., ibid.}  {\bf 103}, 194803 (2009);
 J. Faure, \textit{et al.}, Phys. Plasmas {\bf 17}, 083107 (2010).
\bibitem{Nature} S. P. D. Mangles, \textit{et al., ibid.} \textbf{431}(7008), 535 (2004); C. G. R. Geddes, \textit{et al., ibid.} \textbf{431}(7008), 538 (2004); J. Faure, \textit{et al.}, Nature \textbf{431}(7008), 541 (2004); W. P. Leemans, \textit{et al.}, Nat. Phys. \textbf{2}, 696 (2006);
N. Hafz, \textit{et al.}, Nat. Photon. \textbf{2}, 571 (2008).
\bibitem{LI} S. V. Bulanov, \textit{et al.}, Phys. Rev. Lett. {\bf 91}, 085001 (2003).
\bibitem{LI-exp} M. Kando, \textit{et al.}, Phys. Rev. Lett. \textbf{99}, 135001 (2007); 
M. Kando, \textit{et al., ibid.} \textbf{103}, 235003 (2009).
\bibitem{Liu} M. N. Rosenbluth and C. S. Liu, Phys. Rev. Lett. \textbf{29}, 701 (1972).
\bibitem{Brantov} A. V. Brantov, \textit{et al.}, Phys. Plasmas \textbf{15}, 073111 (2008).
\bibitem{Esirkepov_code} T. Zh. Esirkepov, Comput. Phys. Comm. \textbf{135}, 144 (2001).
\bibitem{others} S. S. Bulanov,  \textit{et al.}, Phys. Rev. E {\bf 73}, 036408 (2006);
V. V. Kulagin, \textit{et al.}, Phys. Plasmas {\bf 14}, 113101 (2007);
D. Habs, \textit{et al.}, Appl. Phys. B, {\bf 93}, 349 (2008);
T. Zh. Esirkepov, \textit{et al.},  Phys. Rev. Lett. {\bf 103}, 025002 (2009);
 H. Wu, \textit{et al., ibid.} {bf 104}, 234801 (2010);
 Ji, L. L., \textit{et al., ibid.} {bf 105}, 025001 (2010).
S. S. Bulanov, \textit{et al.}, Phys. Lett. A, {\bf 374}, 476 (2010).
\bibitem{BellKirk} A. R. Bell and J. G. Kirk, Phys. Rev. Lett. \textbf{101}, 200403 (2008); 
A. M. Fedotov, \textit{et.al.}, Phys. Rev. Lett. \textbf{105}, 080402 (2010);  S. S. Bulanov, \textit{et.al.}, Phys. Rev. Lett. {\bf 105}, 220407, (2010).
\bibitem{H&E} W. Heisenberg and H. Euler, Z. Phys. \textbf{98}, 714 (1936); J. Schwinger, Phys. Rev. \textbf{82}, 664 (1951).
\bibitem{NarozhnyBulanovMurPopov} N. B. Narozhny, \textit{et al.}, Phys. Lett. A \textbf{330}, 1 (2004); S. S. Bulanov, \textit{et al.,} Phys. Rev. Lett. {\bf 104}, 220404, (2010).


\end{thebibliography}
\end{document}